\documentclass[twocolumn,english,aps,prl,reprint,nofootinbib,superscriptaddress]{revtex4-1}
\usepackage[T1]{fontenc}
\usepackage[latin9]{inputenc}
\usepackage{textcomp}
\usepackage{amsmath}
\usepackage{amssymb}
\usepackage{graphicx}
\usepackage{esint}

\makeatletter

\newcommand{\lyxmathsym}[1]{\ifmmode\begingroup\def\b@ld{bold}
  \text{\ifx\math@version\b@ld\bfseries\fi#1}\endgroup\else#1\fi}

\@ifundefined{textcolor}{}
{%
 \definecolor{BLACK}{gray}{0}
 \definecolor{WHITE}{gray}{1}
 \definecolor{RED}{rgb}{1,0,0}
 \definecolor{GREEN}{rgb}{0,1,0}
 \definecolor{BLUE}{rgb}{0,0,1}
 \definecolor{CYAN}{cmyk}{1,0,0,0}
 \definecolor{MAGENTA}{cmyk}{0,1,0,0}
 \definecolor{YELLOW}{cmyk}{0,0,1,0}
}

\usepackage{babel}
\usepackage{float}\usepackage{braket}\usepackage{braket}

\makeatother

\usepackage{babel}
\begin{document}

\author{K. Kolasi\'{n}ski}

\affiliation{AGH University of Science and Technology, Faculty of Physics and
Applied Computer Science,\\
 al. Mickiewicza 30, 30-059 Kraków, Poland}

\affiliation{Institut Néel, CNRS and Université Joseph Fourier, \\
 BP 166, 38042 Grenoble, France}

\author{H. Sellier}

\affiliation{Institut Néel, CNRS and Université Joseph Fourier, \\
 BP 166, 38042 Grenoble, France}

\author{B. Szafran}

\affiliation{AGH University of Science and Technology, Faculty of Physics and
Applied Computer Science,\\
 al. Mickiewicza 30, 30-059 Kraków, Poland}

\title{Conductance measurement of spin-orbit coupling in the two-dimensional
electron systems with in-plane magnetic field}
\begin{abstract}
We consider determination of spin-orbit (SO) coupling constants for
the two-dimensional electron gas from measurements of electric properties
in rotated in-plane magnetic field. 
Due to the SO coupling the electron backscattering is accompanied
by spin precession and spin mixing of the incident and reflected electron
waves. The competition of the external and SO-related magnetic fields
produces a characteristic conductance dependence on the in-plane magnetic
field value and orientation which, in turn, allows for determination
of the absolute value of the effective spin-orbit coupling constant
as well as the ratio of the Rashba and Dresselhaus SO contributions.

\end{abstract}
\maketitle
\emph{Introduction.} Charge carriers in semiconductor devices are
subject to spin-orbit (SO) interactions \cite{Manchon2015} stemming
from the anisotropy of the crystal lattice and/or the device structure.
The SO interactions translate the carrier motion into an effective
magnetic field 
leading to carrier spin relaxation and dephasing \cite{Ohno1999,Dyakonov1971,Kainz2004},
spin Hall effects \cite{Hirsch1999,Sinova2004,Kato2004}, formation
of topological insulators \cite{Koenig2007}, persistent spin helix
states \cite{Bernevig2006,Koralek2009,Walser2012}, Majorana fermions
\cite{Mourik2012}. Moreover, the SO coupling paves the way to spin-active
devices, including spin-filters based on quantum point contacts (QPCs)
\cite{Debray2009} or spin transistors \cite{Datta1990,Schliemann2003,Zutic2004,Chuang2015,Bednarek2008},
which exploit the precession of the electron spin in the effective
magnetic field \cite{Meier}. The most popular playground for studies
of spin effects and construction of spin-active devices is the two-dimensional
electron gas (2DEG) confined at an interface of an asymmetrically
doped III-V heterostructure, with a strong built-in electric fields
in the confinement layer giving rise to the Rashba SO coupling \cite{Bychkov1984}
and with the Dresselhaus coupling due to the anisotropy of the lattice
which is enhanced by a strong localization of the electron gas in
the growth direction \cite{Winkler}.

The SO interaction is sample-dependent and its characterization is
of a basic importance for description of spin-related phenomena and
devices. The SO coupling constant are derived from the Shubnikov-de
Haas \cite{Nitta1997,Engels1997,Lo2002,Kwon2007,Grundler2000,Kim2010,Das1989,Park2013}
oscillations, antilocalization in the magnetotransport \cite{Koga2002},
photocurrents \cite{Ganichev2004}, or precession of optically polarized
electron spins as a function of their drift momentum \cite{Meier2007}.
Usually both the Rashba and Dresselhaus interactions contribute to
the overall SO coupling. Separation of contributions of both types
of SO coupling is challenging and requires procedures based on optical
polarization of the electron spins \cite{Ganichev2004,Meier2007,Wang13}.
In this Letter we investigate the possibility for extraction of the
Rashba and Dresselhaus constants from a purely electric measurement
of the two-terminal conductance. 
The proposed method does not involve application of optical excitation
\cite{Ganichev2004,Meier2007} or a particularly complex gating \cite{Meier2007}.
The procedure given below requires rotation of the sample in an external
in-plane magnetic field \cite{Meckler09}, which is straightforward
as compared to application of the rotated electric field to 2DEG \cite{Meier2007}.
Also, the present approach is suitable for high mobility samples and
goes without analysis of the localization effects in the magnetotransport
\cite{Koga2002}.



The procedure which is proposed below bases on an idea that the effects
of the SO coupling related to the wave vector component in the direction
of the current flow can be excluded by a properly oriented external
in-plane magnetic field. The procedure exploits spin effects of backscattering
-- due to intentionally introduced potentials -- or simply to intrinsic
imperfections within the sample. In particular we show that 
the linear conductance of a disordered sample reveals an oscillatory
behavior as a function of the magnetic field direction and amplitude.
The dependence allows one to determine the strength of the SO interaction
as compared to the spin Zeeman effect as well as the relative strength
of both Rashba and Dresselhaus contributions.

\begin{figure}[h]
\begin{centering}
\includegraphics[width=0.6\columnwidth]{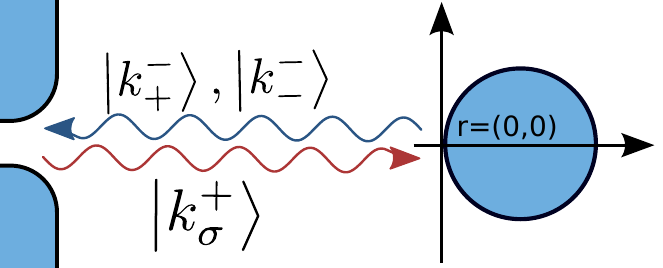} 
\par\end{centering}

\caption{\label{fig:scattering-1}Sketch of considered scattering process.
The electron wave is incoming from the left from a source (a QPC for
instance) in spin state $\sigma$, propagates to the right and is
backscattered at position $r=(0,0)$ by the potential barrier induced
by the impurity.}
\end{figure}

\emph{Spin-dependent scattering model.} Let us start from a simple
model of electron scattering (see Fig. \ref{fig:scattering-1}). The
electron is injected to the system from e.g. a QPC and comes to the
potential defect from the left. The defect is taken as an infinite
potential step, so that the scattering probability is 1. The incident
and backscattered waves are denoted by $\left|k_{\sigma}^{+}\right\rangle $
where $k_{\sigma}^{\pm}$ stands for the absolute value of the wave
vector for the spin state $\sigma$ and the superscript sign indicates
the electron incoming from the left ($+$) or backscattered ($-$).
Only the backscattering which returns the carriers to the QPC can
alter the conductance, so we consider the scattering wave function
along the line between the QPC and the defect 
\begin{equation}
\left|\Psi_{\sigma}\right\rangle =e^{ik_{\sigma}^{+}r}\left|k_{\sigma}^{+}\right\rangle +\Sigma_{\sigma'}a_{\sigma\sigma'}e^{-ik_{\sigma'}^{-}r}\left|k_{\sigma'}^{-}\right\rangle ,\label{eq:scatteq-1}
\end{equation}
where 
$a_{\sigma\sigma'}$ stand for the scattering amplitudes. 

Scattering at other angles does not decrease the conductance and is
neglected for a moment. Within the 2DEG, outside the scattering center
and the QPC channel the 2D electron Hamiltonian for in-plane field
${\bf B}=(B_{\mathrm{x}},B_{\mathrm{y}},0)$ reads 
\begin{equation}
H=E_{\mathrm{kin}}{\bf I}+\sigma_{\mathrm{x}}(\alpha{k}_{\mathrm{y}}-\beta k_{\mathrm{x}}+b_{\mathrm{x}})+\sigma_{\mathrm{y}}(\beta{k}_{\mathrm{y}}-\alpha k_{\mathrm{x}}+b_{\mathrm{y}}),
\end{equation}
where $\boldsymbol{E}_{\mathrm{kin}}=\frac{\hbar^{2}\boldsymbol{k}^{2}}{2m_{\mathrm{eff}}}$,
 $b_{\mathrm{x/y}}=\frac{1}{2}g\mu_{\mathrm{B}}B_{\mathrm{x/y}}$,
$m_{\mathrm{eff}}$ is the electron effective mass, $\alpha$ and
$\beta$ are the Rashba and Dresselhaus constants. The spin Zeeman
effect is introduced via Pauli matrices $\sigma_{\mathrm{x/y}}$ and
the Zeeman energy will be denoted below by $E_{B}=\frac{1}{2}g\mu_{\mathrm{B}}|B|=\sqrt{b_{x}^{2}+b_{y}^{2}}$. 

Note, that we use the symmetric gauge $\boldsymbol{A}=(B_{{y}}z,\lyxmathsym{\textminus}B_{{x}}z,0)$
then by choosing the plane of the 2DEG confinement to be located at
$z=0$, we get $\boldsymbol{A}=\boldsymbol{0}$, and the magnetic
field enters the Hamiltonian only via the spin Zeeman term i.e. the
orbital effects do not affect the electron transport. 

Let us first neglect the Dresselhaus coupling ($\beta=0$). Plane
wave solution for the eigenvalues of the Schrödinger equation gives
\begin{equation}
E_{\mathrm{\sigma}}=\frac{\hbar^{2}\boldsymbol{k}^{2}}{2m_{\mathrm{eff}}}+\sigma\left|\boldsymbol{p}\right|,\label{eq:reldysp-1}
\end{equation}
with $\sigma=\{+,-\}$ denoting projections of the spin on the direction
of polarization $\boldsymbol{p}=\left(\alpha{k}_{\mathrm{y}}+b_{\mathrm{x}},-\alpha k_{\mathrm{x}}+b_{\mathrm{y}}\right)$,
and eigenvectors 
\begin{equation}
\left|k_{\sigma}^{\pm}\right\rangle =\frac{1}{\sqrt{2}}\left(\begin{array}{c}
1\\
\sigma\frac{p_{\mathrm{x}}^{\pm}+ip_{\mathrm{y}}^{\pm}}{p^{\pm}}
\end{array}\right)\equiv\frac{1}{\sqrt{2}}\left(\begin{array}{c}
1\\
\sigma e^{i\phi\left(\boldsymbol{k}_{\sigma}^{\pm},\boldsymbol{B}\right)}
\end{array}\right),\label{eq:modes-1}
\end{equation}
for the incident ($+$) and backscattered ($-$) directions of the
electron motion with $p^{\pm}=|\boldsymbol{p}^{\pm}|$. Due to the
assumed infinite scattering potential, the wave function in Eq. \eqref{eq:scatteq-1}
has to vanish at $r=0$ (see Fig. \ref{fig:scattering-1}), $\Psi_{\sigma}\left(r=0\right)=\left|k_{\sigma}^{+}\right\rangle +\Sigma_{\sigma'}a_{\sigma\sigma'}\left|k_{\sigma'}^{-}\right\rangle =0,$
hence 
\begin{equation}
a_{\sigma\sigma'}=-\sigma'\frac{\sigma e^{i\phi\left(\boldsymbol{k}_{\sigma}^{+},\boldsymbol{B}\right)}+\sigma'e^{i\phi\left(\boldsymbol{k}_{-\sigma'}^{-},\boldsymbol{B}\right)}}{e^{i\phi\left(\boldsymbol{k}_{+}^{-},\boldsymbol{B}\right)}+e^{i\phi\left(\boldsymbol{k}_{-}^{-},\boldsymbol{B}\right)}}.\label{eq:rss-1}
\end{equation}
In the following we use In$_{0.5}$Ga$_{0.5}$As material parameters
with $m=0.0465m_{0}$, Landé factor $g=9$, and the Fermi energy $E_{\mathrm{F}}=20$meV.
For the bulk Rashba \cite{silva} constant $\alpha_{3D}=57.2$ \AA{}$^{2}$,
the 2D value is $\alpha=\alpha_{3D}F_{z}$, where $F_{z}$ is the
electric field in the growth-direction. The Rashba constant can be
controlled by the external voltages \cite{Nitta1997} and for In$_{0.5}$Ga$_{0.5}$As
SO coupling constants of the order of 5 to 10 meV nm \cite{Nitta1997}
were recorded.

\begin{figure}[h]
\begin{centering}
\includegraphics[width=0.8\columnwidth]{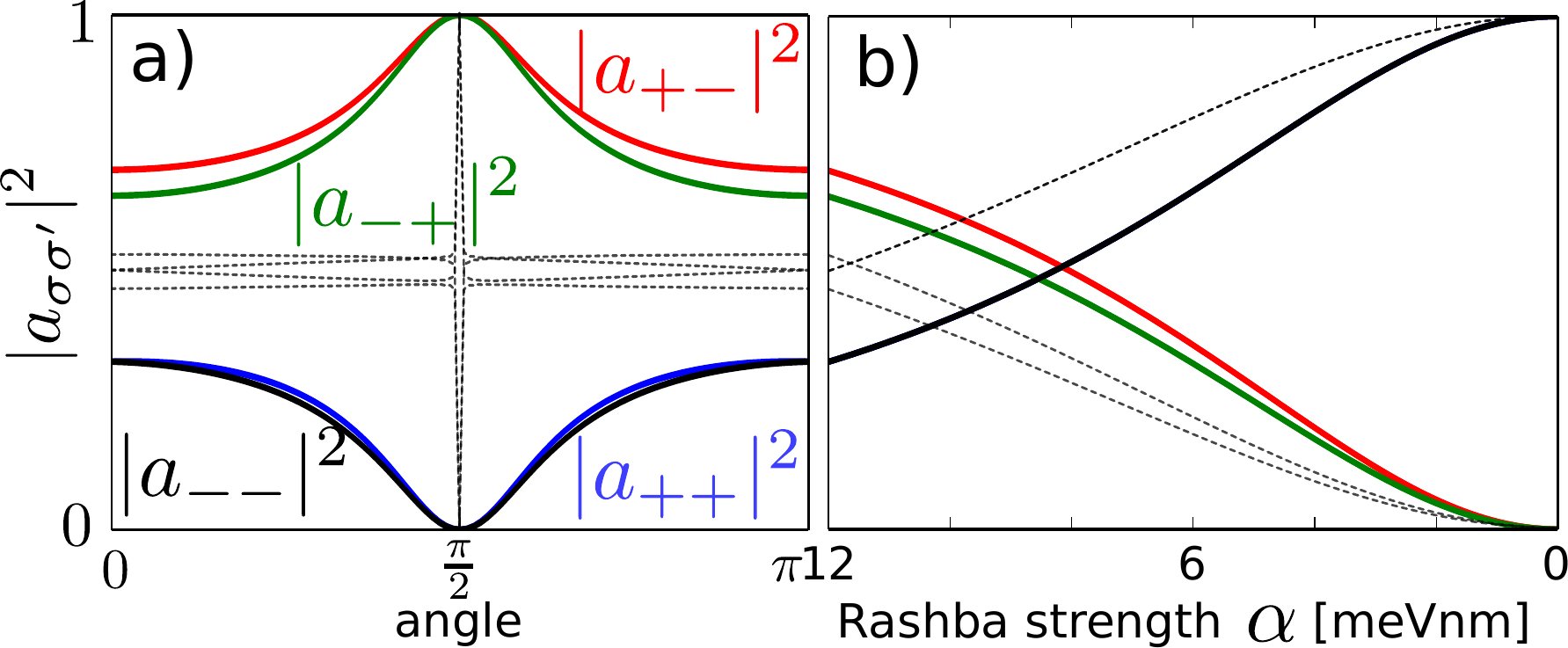} 
\par\end{centering}

\caption{\label{fig:assp-1}(a) Scattering amplitudes calculated from Eq. \eqref{eq:rss-1}
for $\alpha=12$meVnm and $\beta=0$ as a function of angle formed
by the magnetic field vector and the $x$ axis $\theta$. (b) Same
as in (a), but for a fixed angle $\theta=\pi$ as a function of Rashba
constant $\alpha$. In (a) and (b) the solid lines show the result
for $B=5$T and the black dashed lines for $B=7.2$T. }
\end{figure}

In Fig. \ref{fig:assp-1}(a) we present the scattering amplitudes
$a_{\sigma\sigma'}$ obtained from Eq. \eqref{eq:rss-1} as a function
of the direction of the magnetic field $\boldsymbol{B}=\left(B\cos\left(\theta\right),B\sin\left(\theta\right)\right)$,
with $B=5$T for scattering along the $x$ direction, $\boldsymbol{k}=(k_{x},0)$.
Note that for the magnetic field oriented in the $y$ direction $\theta=\pi/2$,
i.e. for $B=\left(0,B_{\mathrm{y}}\right)$ the diagonal elements
of the scattering amplitudes are zero. This is a special case for
which the spinor in Eq. \eqref{eq:modes-1} can be written in form
$\left|k_{\sigma}^{\pm}\right\rangle =\left(\begin{array}{c}
1\\
i\sigma d_{\pm}
\end{array}\right),$ where $d_{\pm}=\mathrm{sign\left(-\alpha k_{\mathrm{x}}^{\pm}+b_{\mathrm{y}}\right)}$.
For a weak magnetic field $\left|\alpha k_{\mathrm{x}}^{\pm}\right|>\left|b_{\mathrm{y}}\right|$,
we get $d_{\pm}=\mp$, and the orthogonality relation $\braket{k_{\sigma'}^{d'}|k_{\sigma}^{d}}=\frac{1}{2}\left(1+\sigma\sigma'dd'\right)$,
gives zero for the backscattering to states with the same spin projection
on the polarization vector (${\bf p}$), $\braket{k_{\sigma}^{-}|k_{\sigma}^{+}}=0$
{[}see Fig. 2(a){]}. On the other hand for high magnetic field $\left|\alpha k_{\mathrm{x}}^{\pm}\right|<\left|b_{\mathrm{y}}\right|$,
we get $d_{\pm}=1$, and the spin projection on the polarization vector
is conserved $\braket{k_{\sigma}^{-}|k_{\sigma}^{+}}=1$.

In Fig. \ref{fig:assp-1}(b) we show the evolution of the scattering
amplitudes for the orientation of magnetic field fixed at $\theta=\pi$,
as a function of the Rashba constant $\alpha$. Then, in Eq. (2) $k_{y}=0$
and $b_{y}=0$. The scattering amplitudes cross near $\alpha\approx8$meVnm
{[}Fig. 2(b){]}. In this point the Zeeman energy $E_{B}$ is equal
to the SO coupling energy $E_{SO}=\alpha k_{x}$. 
For $\alpha k_{x}=E_{B}$ the off-diagonal terms are: $-b_{x}\sigma_{x}+(E_{B}+b_{y})\sigma_{y}$,
for which the scattering amplitudes (5) for eigenvectors (4) simplify
to $\left|a_{\sigma\sigma'}\right|^{2}=\frac{1}{2}$ for any $\sigma\sigma'$
and for any in-plane orientation of \textbf{B} vector. The $E_{B}\approx E_{SO}$
case is presented in Fig. \ref{fig:assp-1}(a) where the black dashed
lines show the scattering amplitudes for $B=7.2$T, which shows that
almost complete spin mixing $\left|a_{\sigma\sigma'}\right|^{2}\approx\frac{1}{2}$
is present for any angle.

Let us now include the Dresselhaus SO coupling. The 2D Dresselhaus
constant is given by $\beta=\beta_{3D}\langle k_{z}^{2}\rangle=\beta_{3D}\frac{\pi^{2}}{d^{2}}$,
where $\beta_{3D}$ is the bulk constant and $d$ is the width of
the 2DEG confinement in the growth direction. We consider $\beta$
values from 0 to $\simeq\alpha$ \cite{Bernevig2006,Koralek2009}.
The cubic Dresselhaus interaction is neglected as a small effect \cite{Koralek2009}.
In the absence of the $B$ field, for the electron incident along
the $x$ direction i.e. $k_{\mathrm{y}}=0$, the polarization direction
is $\boldsymbol{p}=\left(-\beta k_{\mathrm{x}},-\alpha k_{\mathrm{x}}\right)$
with the energy eigenvalues $E_{\sigma}=\frac{\hbar^{2}k_{\mathrm{x}}^{2}}{2m_{\mathrm{eff}}}+\sigma k_{\mathrm{x}}\sqrt{\alpha^{2}+\beta^{2}}$.
As a result the Dresselhaus interaction sets the direction of the
electron spin polarization to $\theta=\arctan\left(\frac{\alpha}{\beta}\right)$
and increases the effective SO coupling constant to $\gamma_{\mathrm{eff}}=\sqrt{\alpha^{2}+\beta^{2}}.$

The above conclusions can also be reached by a direct inspection of
the off-diagonal part of Hamiltonian (2) for the electron transport
along the $x$ direction ($k_{y}=0$, $k_{x}=k_{F}$). The effective
magnetic field in Eq. (2) is $(-\beta k_{F}+b_{x},-\alpha k_{F}+b_{y})$.
Both components of the effective magnetic field vanish for 
\begin{equation}
\tan\theta=\frac{b_{y}}{b_{x}}=\frac{\alpha}{\beta}
\end{equation}
and 
\begin{equation}
E_{\mathrm{B}}=\frac{1}{2}g\mu_{\mathrm{B}}B=\sqrt{b_{x}^{2}+b_{y}^{2}}=k_{F}\gamma_{\mathrm{eff}}\equiv E_{\mathrm{SO}}.
\end{equation}

For illustration we calculated the electron density at the source
position -- including the incident and backscattered waves using Eqs.
(1,5) as $\rho=\sum_{\sigma}\braket{\Psi_{\sigma}|\Psi_{\sigma}}$.
The backscattering probability is roughly proportional to the electron
density at the QPC \cite{Kolasinski2014Slit}. The electron density
-- is depicted in Fig. 3(b-d) for $\alpha=12$meVnm, $\beta=0$; and
Fig. 3(c-d) $\alpha=9$meVnm, $\beta=8$meVnm. These values produce
the same effective coupling constant $\gamma_{\mathrm{eff}}\approx12$meVnm.

\begin{figure}[h]
\begin{centering}
\includegraphics[width=1\columnwidth]{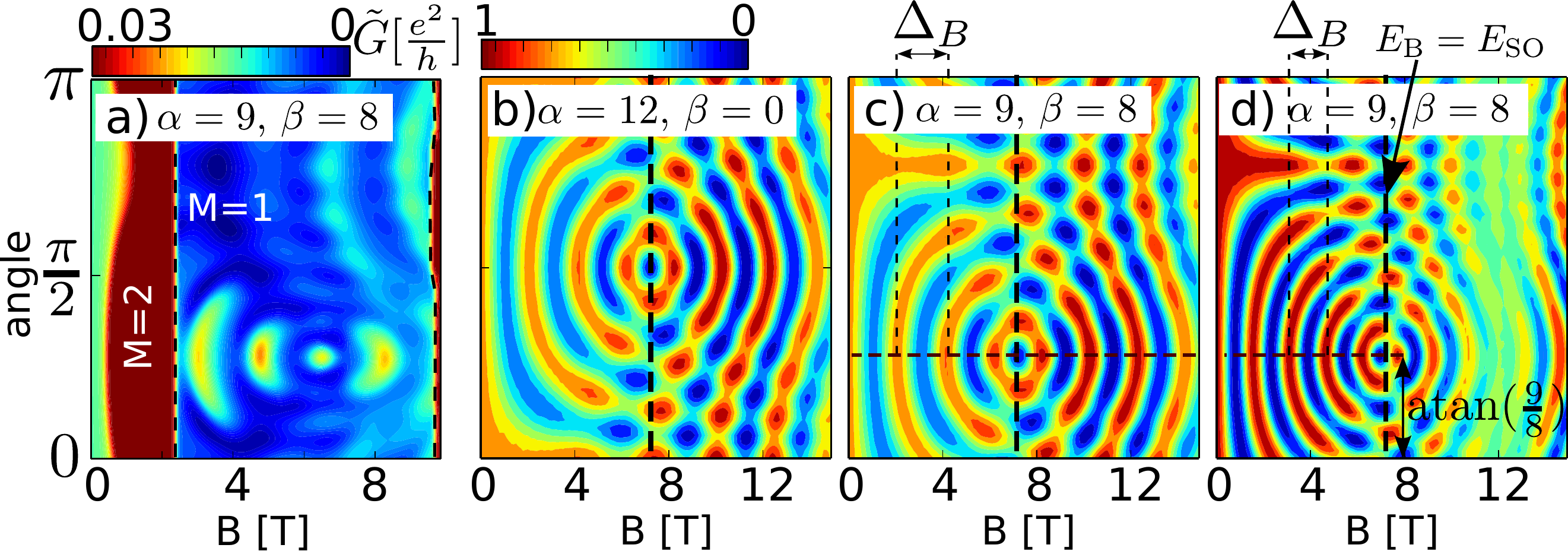} 
\par\end{centering}

\caption{\label{fig:rdx} (a) Reduction of QPC {[}Fig. 1{]} conductance $\tilde{G}=M\frac{\mathrm{e}^{2}}{\mathrm{h}}-G$
from its quantized value for $M$ subbands {[}$E_{F}$=20 meV{]} passing
across the QPC for a potential defect at a distance of 1500 nm from
the QPC as a function of the in-plane magnetic field value and orientation.
The results were calculated numerically within the Landauer approach.
The QPC gate potential modeled with analytical formulas for a rectangle
gate adapted from Ref. \cite{Davies1995}. (b) Charge density at the
entrance to the QPC calculated with a simple model of Eq.(5) for the
QPC at 1500 {nm} from the scatterer. (c) Same as (b) but with Rashba
and Dresselhaus SO interactions present. (d) Same as (c) but with
the source at 2000{nm} from the scatterer. The values of the coupling
constants $\alpha$ and $\beta$ are given in meV$\times$nm units.
The vertical dashed line in(b-d) indicates $B=7.2$ T for which the
Zeeman energy is equal to the SO coupling energy, $E_{B}=E_{SO}$,
see text. The horizontal dashed line shows the angle $\arctan\frac{\alpha}{\beta}$ }
\end{figure}

The results of Fig. 3(b-d) contain a distinct circular pattern in
the $\theta,B$ plane. The position of the center is given by Eqs.
(6,7). The angular coordinate of the center allows one to determine
the ratio of the Rashba and Dresselhaus constants and the SO coupling
constant $\gamma_{\mathrm{eff}}$ can be read-out from the position
of the center of the pattern on the $B$ scale, provided that the
Fermi wave vector is known. In presence of SO coupling and / or the
Zeeman effect $k_{F}$ is spin dependent \cite{Winkler}. However,
for the $E_{B}=E_{SO}$ the off-diagonal terms of the Hamiltonian
(2) vanish and the Fermi wavevector is directly related to the Fermi
energy $E_{F}=\frac{\hbar^{2}k_{F}^{2}}{2m_{\mathrm{eff}}}$, which
for the adopted parameters gives $k_{F}=0.156$/nm. For $\gamma_{\mathrm{eff}}=12$
meV nm one obtains $E_{SO}=1.875$ meV, which coincides with $E_{B}$
for $B=7.2$ T (see Fig. 3(b-d)).

In Fig. \ref{fig:rdx}(c) one notices a reduction of the period with
respect to \ref{fig:rdx}(b) with the source-impurity distance increased
to 2000 nm from 1500 nm. The period of the oscillations $\Delta_{\mathbb{\mathrm{B}}}$
is $\Delta_{\mathbb{\mathrm{B}}}^{\left(c\right)}\approx1.5$T in
(c), and $\Delta_{\mathbb{\mathrm{B}}}^{\left(b\right)}\approx2.0$T
in (b). The ratio $\Delta_{\mathbb{\mathrm{B}}}^{\left(c\right)}/\Delta_{\mathbb{\mathrm{B}}}^{\left(b\right)}\approx3/4$,
is exactly an inverse of the source-impurity $d_{\mathrm{s-i}}$ distance
ratio. 



\emph{Coherent quantum transport calculations.} With the intuitions
gained by the simple analytical model we can pass to the calculations
of the coherent transport using a standard numerical method \cite{Kolasinski2016Lande},
based on the quantum transmitting boundary solution of the quantum
scattering equation at the Fermi level implemented in the finite difference
approach, which produces the electron transfer probability used in
the Landauer formula for conductance summed over the subbands of the
channels far from the scattering area. Zero temperature is assumed.
For the numerical calculations we consider a channel extended along
the $x$ direction, hence $k_{x}$ in Eq. (2) remains a quantum number
characterizing the asymptotic states of the channel. Within the computational
box the wave vector is replaced by an operator ${\bf k}=(k_{x},k_{y})=-i\nabla$.

We consider a QPC/defect system of Fig. 1. The results presented in
Fig. 3(a) indicate a reduction of the conductance below the maximal
value $M\frac{\mathrm{e}^{2}}{h}$ for $M$ subbands passing across
the QPC. The central position of the pattern nearly coincides with
the one of Fig. 3(c). The local extremum of conductance in the center
of the pattern indicates the value and orientation of the external
magnetic field which lifts the SO interaction effects. The angular
position of Fig. 3(c) is exactly reproduced, and the amplitude of
the field is $B=6.5$ T instead of $B=7.2$ T. The deviation in the
location of the central point in the $B$ axis in Fig. 3(a) results
from the confinement in the QPC channel which is not included in our
free particle model (see below).



The effects described so far dealt with interference of the electron
waves between the source (QPC) and the defect. In fact, the role of
the source can be played by any scattering center, and the extraction
of the SO coupling constant requires a presence of two or more scatterers
to allow for formation of standing waves described in the previous
section. For the rest of the paper we consider a channel of homogenous
width $W$, which carries $M$ transport modes at the Fermi level.
In Fig. 4(d) we presented the conductance results for a clean channel
of width $W=180$nm and the computational box of length $L=1.6\mu$m.
A smooth potential barrier is introduced across the channel with height
10 meV and width 200nm. Depending on the orientation of the magnetic
field the number of transport modes varies between $M=17$ and $M=18$.
The simulation was performed for $\alpha=9$meVnm and $\beta=8$meVnm
as in Fig. 3(b,c). The conductance plot possesses an extremum precisely
at the angle of $\theta=\arctan\frac{9}{8}$. The magnetic field of
the extremum is slightly shifted to lower values than 7.2 T -- which
is a result of the reduction of $k_{x}$ within the potential barrier.
The lack of conductance oscillations that were observed above in Fig.
3 results from a small barrier length ($d_{s-i}$=200nm).

\begin{figure}[h]
\begin{centering}
\includegraphics[width=1\columnwidth]{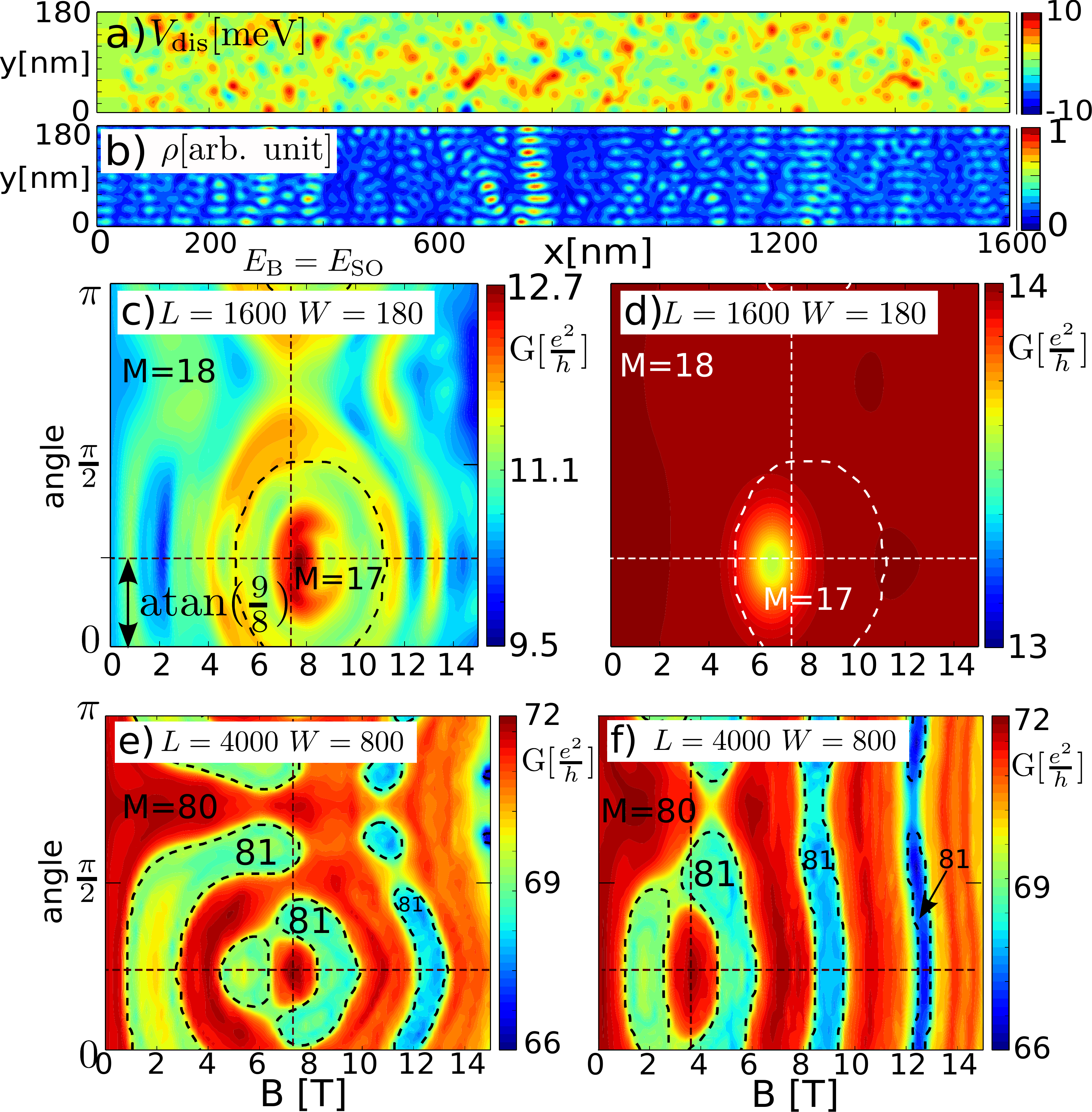} 
\par\end{centering}

\caption{\label{fig:dw}a) Potential disorder in simulated quantum wire. b)
Local density of states obtained for channel from (a) at B=0. c) Conductance
through the wire in (a) as a function of magnetic field amplitude
$B$ and direction \emph{angle}. d) Same as c) but with one potential
barrier in the middle of channel instead of random disorder. e) Same
as (c) but for the wire with length $L=4000$nm and $W=800$nm. f)
Same as (e) but for SO couplings twice smaller $\gamma_{\mathrm{eff}}=6$meVnm.
For comparison of (c,e) see Fig. \ref{fig:rdx}(b). The values $M$
show the number of non-degenerated modes in the channel. }
\end{figure}

The oscillations reappear when one replaces the barrier by a random
disorder due to the random nonmagnetic and spin-diagonal potential
fluctuations. The fluctuations simulate inhomogeneity of the doping
of the potential barrier which provides the charge to the 2DEG. In
2DEG in III-V's due to the spatial separation of the impurities of
the 2DEG, the defects do not introduce any significant contribution
to the spin-orbit interaction (see Ref. \cite{Sellier} and the Supplement
\cite{Supplement}). Figure 4(c) displays the conductance for the
channel of the same width and length. The potential -- displayed in
Fig. 4(a) is locally varied within the range of (-$0.5E_{\mathrm{F}}$,$0.5E_{\mathrm{F}}$).
The perturbation induces a multitude of scattering evens -- the local
density of states at the Fermi level for $B=0$ is displayed in Fig.
\ref{fig:dw}(b). In spite of the complexity of the density of states
the angular shift is still $\arctan\left(9/8\right)\approx\pi/4$.
The shift of the $G$ extremum along the $B$ scale with respect to
$7.2$ T is detectable -- but small and of an opposite sign than in
Fig. 4(d). This shift is related to the fact that for a finite width
channel $k_{y}$ is an operator that mixes the subbands. The wave
vector $k_{y}$ is a well-defined quantum number for electrons moving
in an unconfined space. The small -- but detectable -- effects of
a finite $W$ disappear completely for a wider channel -- which is
illustrated in Fig. 4(e) for $W=0.8$ $\mu$m. Here, the number of
conducting bands varies between 80 and 81. The local extremum of conductance
appears exactly at the positions indicated in the previous section.
Note, that although the number of subbands change by 1 in Fig. 4(c,e),
the variation of conductance is as large as $\sim3e^{2}/h$ in Fig.
4(c) and $\sim6e^{2}/h$ in Fig. 4(e). The conductance variation in
Fig. 3(a) was very small -- since the defect was far away from the
QPC, for the disordered channel it is no longer the case. For completeness
in Fig. \ref{fig:dw}(f) we presented calculations for a twice smaller
SO coupling constants $\alpha=4.5$ meVnm, $\beta=4$ meVnm, and $\gamma_{\mathrm{eff}}=6$meVnm.
The position of the maximal $G$ along the $B$ scale is consistently
reduced from 7.2 T to 3.6 T, and the orientation of the magnetic field
vector corresponding to the extremum is unchanged.

\emph{Summary.} We have shown that the in-plane magnetic field can
lift the off-diagonal terms of the transport Hamiltonian for the two-dimensional
electron gas that result from the Zeeman effect and the SO interaction.
The effect appears only for a value and orientation of the external
magnetic field which excludes the spin mixing effects that accompany
the backscattering in presence of the SO coupling. In consequence
the conductance maps for a system containing two or more scatterers
-- intentionally introduced -- or inherently present in a disordered
sample exhibit a pronounced extremum as a function of the magnetic
field modulus $B$ and orientation $\theta$. An experimental value
of $B$ -- for which the Zeeman energy is equal to the SO coupling
energy -- should allow one to extract the effective SO coupling constant
including both the Rashba and Dresselhaus terms, and the orientation
field indicates the relative contributions of both. The results indicate
the ratio of the Dresselhaus and Rashba constants is exactly resolved
by the procedure, and the amplitude of the magnetic field -- hence
the effective SO constant varies only within a 10\% from the exact
value depending on the channel width and disorder profile.

\textit{Acknowledgments} This work was supported by National Science
Centre (NCN) grant DEC-2015/17/N/ST3/02266, and by PL-Grid Infrastructure.
The first author is supported by the Smoluchowski scholarship from
the KNOW funding and by the NCN Etiuda stipend DEC-2015/16/T/ST3/00310.

 \bibliographystyle{apsrev4-1-nourl}
\bibliography{referencje}

\end{document}


\author{K. Kolasi\'{n}ski}

\affiliation{AGH University of Science and Technology, Faculty of Physics and
Applied Computer Science,\\
 al. Mickiewicza 30, 30-059 Kraków, Poland}

\affiliation{Institut Néel, CNRS and Université Joseph Fourier, \\
 BP 166, 38042 Grenoble, France}

\author{H. Sellier}

\affiliation{Institut Néel, CNRS and Université Joseph Fourier, \\
 BP 166, 38042 Grenoble, France}

\author{B. Szafran}

\affiliation{AGH University of Science and Technology, Faculty of Physics and
Applied Computer Science,\\
 al. Mickiewicza 30, 30-059 Kraków, Poland}

\title{Supplemental Material for Conductance measurement of spin-orbit coupling
in the two-dimensional electron systems with in-plane magnetic field}

\maketitle

\section{Stability of the conductance pattern }

\global\long\def\thefigure{S\arabic{figure}}

\begin{figure}
\centering{}\includegraphics[width=1\columnwidth]{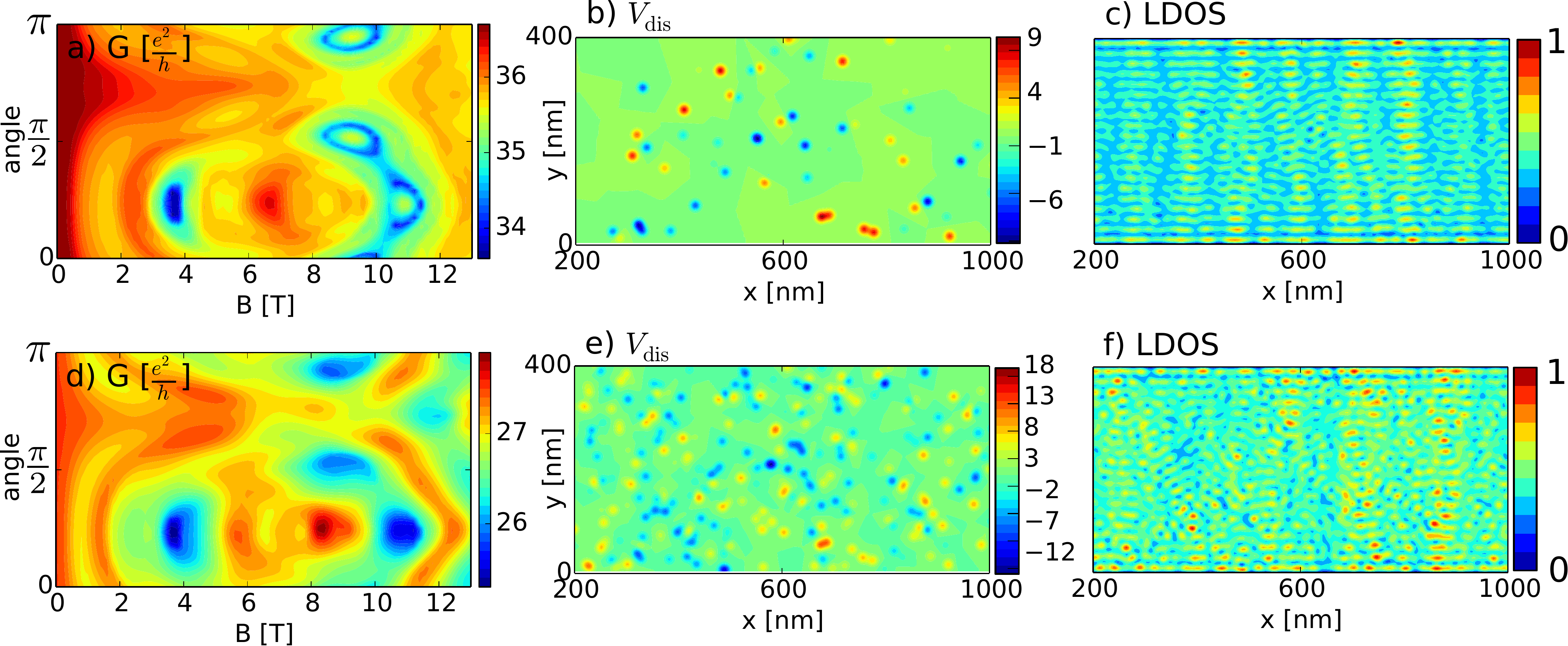}\caption{\label{fig:rc}First column -- conductance as a function of magnetic
field amplitude and angle. Second column -- the random disorder potential
energy in the channel in units of meV. Third column -- resulting LDOS
for $B=0$ T in arbitrary units. }
\end{figure}

The Letter indicated a well resolved extremum of conductance for the
conditions given by Eq. (6) and (7), for which the external magnetic
field cancels the SO effective magnetic field that depends on the
electron wave vector. The presented analysis is based on the electron
transport along the $x$ direction. The conductance extremum was found
for a quantum point contact with a single defect {[}Fig. 3(a){]} and
for a potential barrier {[}Fig. 3(d){]}. These two systems corresponded
to a nearly 1D transport, since the potential barrier was introduced
in a separable manner, and in the quantum point contact an appearance
of the non-zero $k_{y}$ implies removal of the electron from the
beam of backscattered electrons. However, for the perturbed sample
{[}Fig. 4(c,e,f){]} the electrons undergo multiple scattering and
acquire non-zero $k_{y}$ values between the scattering events. Results
of Fig. 4(c,e,f) obtained for a disorder covering basically the entire
width of the sample indicate that the effects of temporarily non-zero
$k_{y}$ for the spin precession averages out to zero and the conductance
extremum stays in the position defined by Eqs. (6,7). In order to
support the discussion further we present in this document the case
with a smaller number of defects, where the averaging may not be complete.
We take a channel of width $W=$400nm and length $L=$1200nm. The
simulation was performed for $\alpha$ = 9meVnm and $\beta$ = 8meVnm
as in Fig. 3(b,c). In the first column of Fig. \ref{fig:rc} we show
the obtained conductance images for different configurations of disorder
potential $V_{\mathrm{dis}}$ in the channel (second column). The
third column present an image of LDOS computed for $B=0$.

One notices the angle shift $\theta=\arctan\left(\frac{\alpha}{\beta}\right)$,
equal to the ratio between Rashba and Dresselhaus strengths, remains
constant for different densities and configurations of disorder. Hence
still for a limited number of defects the measurement of $\theta$
should provide good information about the $\frac{\alpha}{\beta}$.
However, the position of the vertical line for which $E_{B}=E_{SO}$
can be estimated roughly from the obtained images to be between 7
and 8 T which agrees with the assumed value of the SO interaction
in simulated device. The exact position of the center on the $B$
scale for the adopted parameters is 7.2 T. 

To conclude, in general we have found that detecting the angle shift
$\theta$ is more robust against the sample variations than finding
the magnetic field strength symmetry point.

\section{The effect of the impurities on SO coupling constant}

The disorder considered in the Letter is a potential fluctuation,
which results from e.g. non-ideally homogenous doping of the barrier
layer under which the two-dimensional electron gas is confined. In
the 2DEG heterostructures the electron gas is separated from the impurities
by an undoped spacer of a few to about 20 nm. The potential fluctuation
affects the electric field in the growth direction, so an estimation
of the resulting local change of the Rashba SO coupling constant is
needed.

The Rashba SO interaction strength $\alpha=\alpha_{3D}F'_{z}$ results
from the structural inversion asymmetry in the 2DEG confinement plane
and is proportional to the electric field in the growth direction
$F'_{z}$, where for the bulk In$_{0.5}$Ga$_{0.5}$As alloy Rashba
\cite{silva} constant is $\alpha_{3D}=57.2$ \AA{}$^{2}$. The strength
of the electric gradient $F'_{z}$ can be tuned by external voltages
\cite{Nitta1997} and may be varied from 50 to 100 kV/cm in In$_{0.5}$Ga$_{0.5}$As
as reported in Ref. \cite{Nitta1997}. The value of the electric field
can be also changed by random impurities in the structure or result
from the inhomogeneity of the distribution of the impurities, which
may locally change the value of $F'_{z}\rightarrow F'_{z}+\Delta F_{z}$.
The vertical position of the 2DEG sheet is set at $z=0$. In order
to estimate the strength of the electric gradient induced by single
impurity we computed the electric field 
\begin{equation}
\Delta F_{\mathrm{z}}(x,y)=\left.-\frac{\partial}{\partial z}\left(\frac{e}{4\pi\epsilon\epsilon^{*}\left|\boldsymbol{r}_{\mathrm{imp}}-\boldsymbol{r}'\right|}\right)\right|_{\boldsymbol{r}'=(x,y,0)}\label{eq:Fz}
\end{equation}
induced by negatively charged impurity located 10nm from the 2DEG
surface (see Fig. \ref{fig:dft}(c)). The correction to the electric
field due to a single impurity located is negligibly weak as compared
to the nominal value of $F_{\mathrm{z}}'$, hence it can be neglected
in our simulations. 


\section{Friedel oscillations and the effective potential}

In strongly perturbed systems the charge density forms ripples with
the periodicity of half of the Fermi wavelength. We considered the
effect for the present perturbation -- an impurity at a distance from
the 2DEG. We simulated the effect of the single negatively charged
impurity on the electron density using the DFT method described in
Ref. \cite{Kolasinski2016Lande}. In simulations we put $\alpha=10$meVnm
and $\beta=0$ i.e. only the Rashba term is active. The impurity is
introduced 10nm from the the 2DEG computational sheet in the middle
of the computational box. The average electron density in the simulation
is $1\times10^{12}[\frac{1}{\mathrm{cm^{2}}}]$. The self-consistent
DFT solution is reached for temperature $T=1$K. The rest of the parameters
is the same as in the transport simulations.

\begin{figure}
\begin{centering}
\includegraphics[width=1\columnwidth]{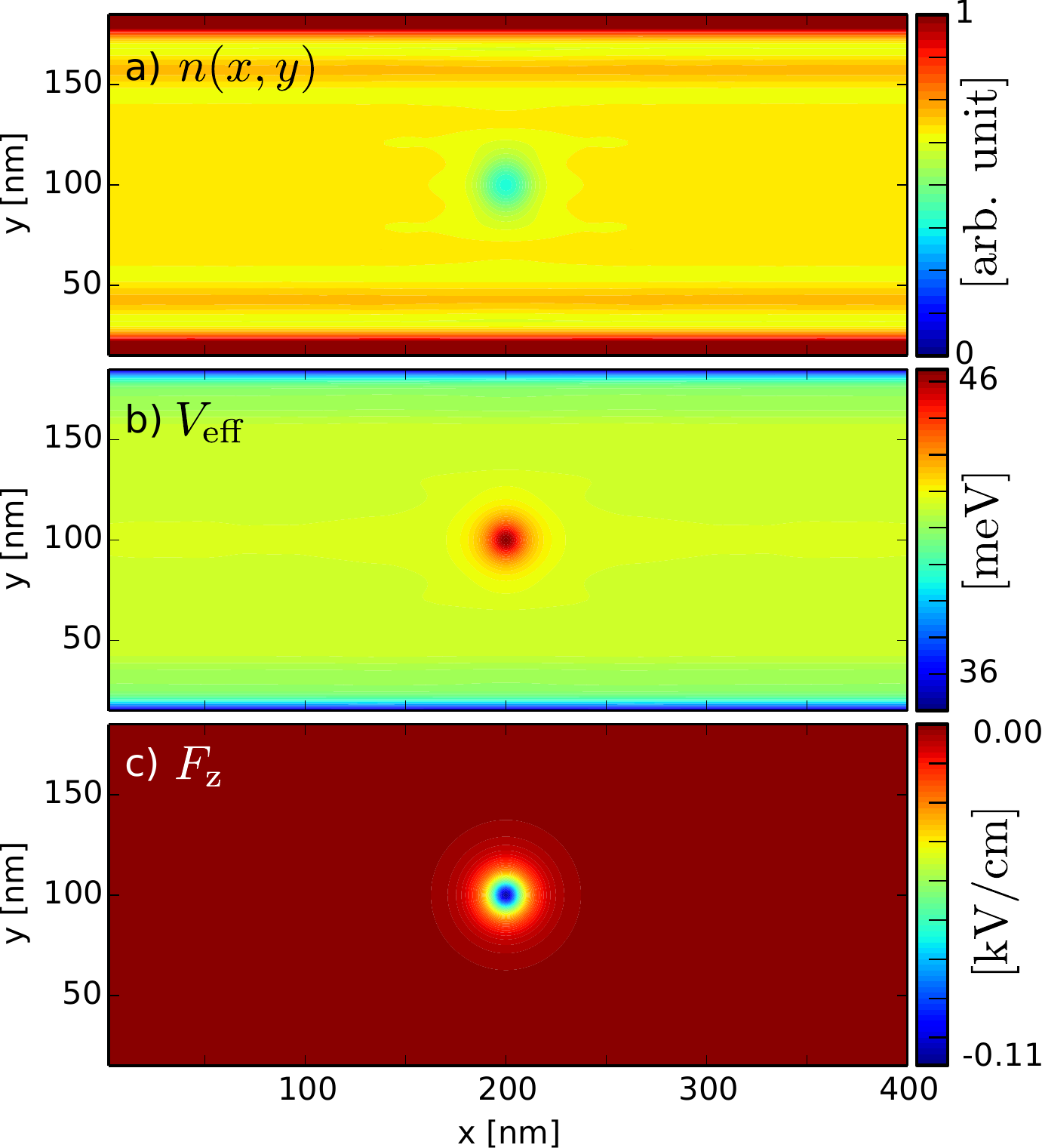} 
\par\end{centering}

\centering{}\caption{\label{fig:dft} (a) The self consistent electron density in the channel
of width 200nm and length 400nm. (b) DFT self-consistent potential
energy. (c) The bare electric field $\Delta F_{\mathrm{z}}(x,y)$
at $z=0$ induced by the impurity (see Eq. \eqref{eq:Fz}).}
\end{figure}

In Fig. \ref{fig:dft}(a) we show the self consistent electron density
obtained from our DFT solver. One may see that the effect of introduced
impurity is relatively weak for a given configuration. The electron
density for $E_{F}=20$ meV is large, hence the screening is quite
efficient, resulting in small depletion in the electron density. The
change of the effective potential energy due to the presence of the
defect is well visible (see Fig. \ref{fig:dft}(b)) and vary from
about 42meV to 46meV.

\begin{figure}
\includegraphics[width=1\columnwidth]{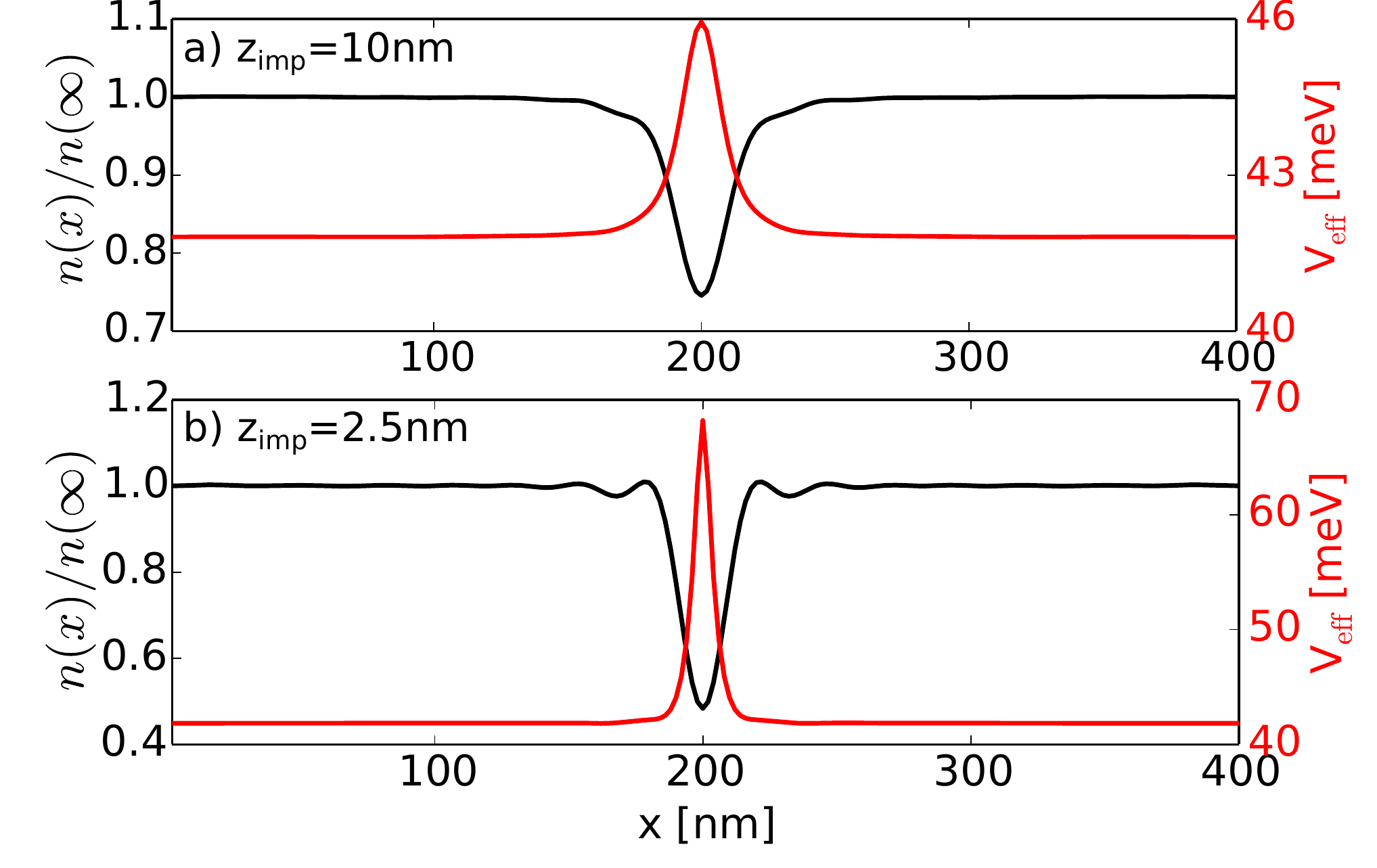}

\centering{}\caption{\label{fig:dft-cs} (a) The cross section of normalized electron density
for $y=100$ nm from Fig. \ref{fig:dft}(a) (black line) and the effective
potential -- potential of the impurity screened by the deformation
of the electron gas (red line). (b) Same as (a) only for the impurity
at 2.5 nm from the electron gas.}
\end{figure}

In Fig. \ref{fig:dft-cs}(a) we plot the cross section of the electron
density and effective potential from Fig. \ref{fig:dft}(a-b) with
peak in the $V_{\mathrm{eff}}$ and deep in electron density $n$
at $x=200$nm. Due to the relatively weak perturbation of the impurity
($\sim25\%$ of depletion) the amplitude of the Friedel oscillations
are very small but still visible in the electron density plot. However,
due to the smearing property of the Coulomb kernel $V(\boldsymbol{r})=\int\boldsymbol{dr'}n(\boldsymbol{r'})/\left|\boldsymbol{r}-\boldsymbol{r}'\right|$
the weak Friedel oscillations in $n$ become even weaker in the resulting
potential energy, hence the lack of the Friedel oscillation in the
obtained $V_{\mathrm{eff}}$ plot. In order to enhance the oscillation
of the charge density in Fig. \ref{fig:dft-cs}(b) we put the impurity
at 2.5nm only from the electron gas. Now, the ripples on the density
are more pronounced, but their are not resolved in the self-consistent
potential anyway. To conclude, the effect of Friedel oscillation on
the DFT potential for the considered perturbation is weak, and the
modification of the charge density by the rotation of the magnetic
field in presence of the SO coupling should not alter the results
for the electron transport.

 \bibliographystyle{apsrev4-1-nourl}
\bibliography{referencje}